# Nursery function rehabilitation projects in port areas can support fish populations but they remain less effective than ensuring compliance to fisheries management

Etienne Joubert[1], Charlotte Sève[2], Stéphanie Mahévas[3], Adrian Bach[1], Marc Bouchoucha[1]

1. Ifremer, Laboratoire Environnement Ressources Provence Azur Corse, CS 20330, F-83507 La Seyne Sur Mer, France

2. Ifremer, Laboratoire Ecologie et Modèles pour l'Halieutique, BP 21105 - 44311 Nantes, France

3. Ifremer, Laboratoire Halieutique Méditerranée, CS 30171 - 34203 Sète Cedex, France

* Corresponding author: etiennejoubert29@gmail.com; +33649293669; Ifremer – Seyne sur Mer, Zone Portuaire de Brégaillon, 83500 La Seyne-sur-Mer

**Abstract:**

1- Conservation measures are implemented to support biodiversity in areas that are degraded or under anthropogenic pressure. Over the past decade, numerous projects aimed at rehabilitating a fish nursery function in ports, through the installation of artificial structures, have emerged. While studies conducted on these solutions seem promising on a very local scale (*e.g.*, higher densities of juvenile fish on artificial fish nurseries compared to bare port infrastructures), no evaluation has been undertaken yet to establish their contribution to the renewal of coastal fish populations or their performance compared to other conservation measures such as fishing regulation.

2- Here, we used a coupled model of fish population dynamics and fisheries management, ISIS-fish, to describe the coastal commercial fish population, the white seabream (*Diplodus sargus*) in the highly artificialized Bay of Toulon.

3- Using ISIS-Fish, we simulated rehabilitation and fisheries management scenarios. We provided the first quantitative assessment of the implementation of artificial structures in ports covering 10% and 100% of the available port area and compared, at population level and fishing fleets level, the quantitative consequences of these rehabilitation measures with fishing control measures leading to strict compliance with minimum catch sizes.

4- The rehabilitation of the nursery function in ports demonstrated a potential to enhance the renewal of fish populations and catches. When the size of projects is small the outcomes they provide remain relatively modest in contrast to the impact of regulatory fishing measures. However, we have demonstrated that combining fishing reduction measures and rehabilitation projects has a synergistic effect on fish populations, resulting in increased populations and catches.

5- *Synthesis and applications*. This study is the first quantitative assessment of fish nursery rehabilitation projects in port areas, by evaluating their effectiveness in renewing coastal fish populations and fisheries and comparing their outcomes with fishing control measures. Small-scale port-area nursery rehabilitation projects can support fish populations, but are less effective than controlling fisheries.

# Introduction

In the context of climate change, human activities have led to a significant decline in marine biodiversity (De Vos et al., 2015; Pimm et al., 2014) and fish stocks (Yan et al., 2021). The two main drivers of this change are overfishing and habitat loss (Yan et al., 2021). This is particularly true in coastal areas where sources of pressure accumulate: the pressure of recreational fishing adds to that of professional fishing. Coastal urbanization processes, and particularly the development of port areas (Meinesz et al., 1991), have led to the destruction and transformation of fish habitats (Mooser et al., 2021; Poursanidis et al., 2018). However, coastal habitats are critical assets for the sustainability of fish stocks, given their indispensable role as nursery grounds. Indeed, these areas can host high concentrations of juvenile fish, enhance juvenile growth with high food availability, decrease mortality due to predation, and actively participate in renewing adult populations (Beck et al., 2001). The survival of fish juveniles in nursery areas is influenced by density-dependent factors, such as the carrying capacity of the habitat, predation, and food availability (Le Pape et al., 2020; Planes et al., 1998) determining juvenile mortality and mainly conditioned by the structural complexity of the habitat (Connell and Jones, 1991; Scharf et al., 2006). Therefore, the quality and the surface area of nurseries are pivotal, determining the influx of new recruits into adult populations each year (Wilson et al., 2016). The increasing degradation of nursery habitats, combined with rising fishing pressure, amplifies the risk of population extinction, thereby posing significant ecological and food security issues (Yan et al., 2021). Considering this, it has become imperative for managers to proactively implement solutions to mitigate these detrimental effects.

To address the decline in fish stocks, European legislation has implemented regulations such as the establishment of a minimum catch size for both commercial and recreational fishing (Cardinale et al., 2017). Nevertheless, the complex socio-economic situation in the Mediterranean Sea sometimes hinders the efficacy of management governance and restrictions are seldom applied in practice

(Cardinale et al., 2017), particularly by recreational fishing with high rates of fish caught below the legal size limit (Font and Lloret, 2014).

More recently, efforts have been underway to mitigate the impact of coastal artificialization on fish populations. They generally involve the implementation of rehabilitation actions (Gann et al., 2019) to fulfil crucial ecosystem functions. To this end, the structural complexity of man-made infrastructures is increasingly taken into account to mimic nature, either during their construction, with specific designs, or post-construction, with the addition of micro-structures (Airoldi et al., 2021). Among these projects, small artificial fish nurseries (1 $m^2$ to 50 $m^2$) are increasingly employed in ports in the Mediterranean with the goal to rehabilitate the fish nursery function(Bouchoucha et al., 2016; Joubert et al., 2023; Patranella et al., 2017). Currently, over thirty ports in the French Mediterranean region are equipped with these structures.

Although monitoring these structures has revealed their effectiveness in increasing local juvenile fish abundance (Bouchoucha et al., 2016; Joubert et al., 2023; Patranella et al., 2017), existing studies are confined to small-scale projects, and surveys exclusively concentrate on the evaluation of fish abundance within ports. The connectivity between artificial fish nurseries and adult populations remains unexplored and the real capacity of these initiatives to support the renewal of adult fish populations with quantitative data has never been assessed (Macura et al., 2019). Currently, this lack of investigation hinders the determination of the effective contribution of these projects in truly rehabilitating the nursery function within ports. The recurring issue of insufficient insight into the true large-scale ecological benefits of marine environmental restoration projects is often inadequately addressed or entirely overlooked in many studies (Airoldi et al., 2021). In a world where financial resources are constrained and decisions regarding interventions must be prioritized, quantifying this contribution would not only aid stakeholders to compare the potential effectiveness of various measures more efficiently (Ward et al., 2022) but also to dimension projects.

To guide stakeholders in the implementation of the most effective conservation measures, it is crucial to anticipate the ecological outcomes and their impact on catches resulting from the management strategies employed (Airoldi et al., 2021). Here, our objective is to dress a first quantitative assessment the tangible impact of establishing artificial nurseries in ports on fish stocks and to provide a quantitative comparison between different restoration and fishing management measures.

Spatial models are widely used for prediction and decision-making in ecology (DeAngelis and Diaz, 2019), especially to predict intricate interactions within ecosystems, whether focusing on population dynamics, anthropogenic activities (Glaum et al., 2020; Plagányi et al., 2014) or conservation strategies (Bach et al., 2022; Gernez et al., 2023). They enable assessing the effectiveness of various scenarios representing adjustments to a reference state to model the actions implemented by management strategies (Refsgaard et al., 2007). Hence, spatial models are increasingly used to evaluate the effectiveness of restoration or rehabilitation measures, by comparing their impact with that of other management strategies (Possingham et al., 2015).

In this study, we used the multi-fleet spatially explicit ISIS-fish model (Mahévas and Pelletier, 2004) to evaluate the impact of four management scenarios on fish stock size, biomass and catches. First, we tested the consequences of implementing artificial structures in ports, covering 10% of the port area, as a potentially attainable objective in the short term, and the hypothetical case of 100% of the available port area equipped with rehabilitation solutions (scenarios 1 and 2, respectively). Then, we studied the impact of applying the strict application of existing regulations in terms of minimum catches (scenario 3). Finally, we evaluated the maximal expected impact scenario of implementing both artificial structures in ports and a strict catch regulation (scenario 4, resulting in the combination of scenarios 1 and 3).

## Material and Methods

### Case study

The study area encompasses the coastal zone near the city of Toulon (France) between cape Sicié to the east (43.045975°E; 5.859140°N) and cape Blanc to the west (43.091305°E; 6.371856°N) (Fig.1), in the southeastern part of the Mediterranean coast. Spanning a surface area of 532 km², it covers all the coastal marine areas that are less than 50 meters deep (Fig. 1). The Bay of Toulon has a mostly man-made coastline, making it suitable for nursery function rehabilitation measures. Anthropogenic activities mainly include tourism and local professional and recreational coastal fishing. The eastern part of the study area falls within the Port Cros National Park (MPA) (Fig. 1) that includes a controlled zone for marine activities and a reinforced protection zone where only professional fishing is allowed (Fig. 1). This study area therefore offers a number of opportunities for implementing management measures.

Some of these coastal areas possess physical characteristics of nurseries for various marine fish species: they feature shallow habitats (0-3m) protected from prevailing winds and waves, with gently sloping bottoms covered with sand, pebbles, cobbles, or algae-covered rocks (Harmelin-Vivien et al., 1995). Based on these characteristics, we estimate the linear length of natural nurseries at 55 820m, at a scale of 1:500 (see Section SI.1 in Supporting Information for further details). During benthic settlement, a proportion of juvenile fish enter port areas (Bouchoucha et al., 2016; Joubert et al., 2023) and can be hosted in infrastructures at less than 2m depth (docks). Thus, we estimate the area of port nurseries to cover 24 771m (Fig. 2).

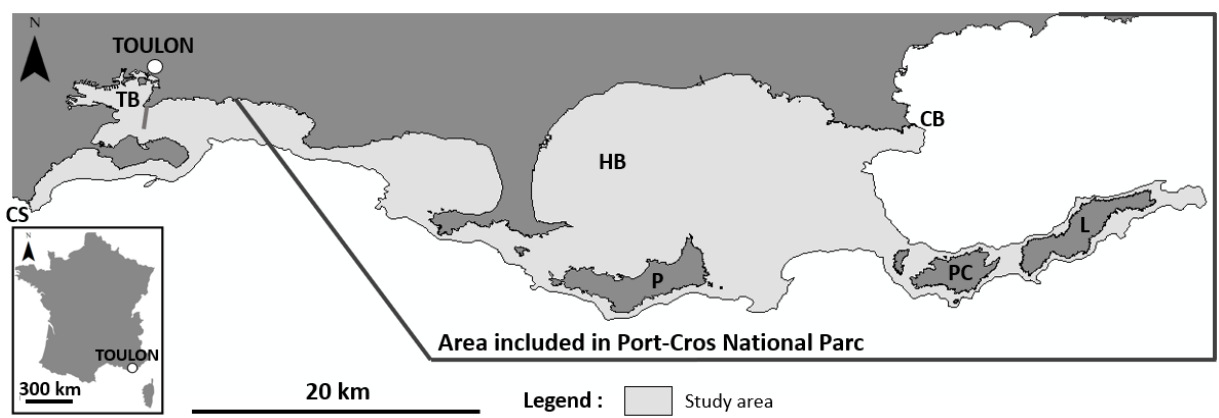

*Figure 1 : Map of the study area, TB:Toulon Bay, HB: Hyères Bay, P: Porquerolles, PC: Port-Cros, L: Levant Island, CS: Cape Sicié, CB: Cape Blanc.*

We focussed on the white seabream (*Diplodus sargus sargus*, Linnaeus, 1758, hereafter *D. sargus*). This coastal, nursery-dependent fish species inhabits areas at depths ranging from 0 to 50m (Harmelin-Vivien et al., 1995) and is of significant economic importance for artisanal and recreational fishers

(Vigliola et al., 1998). Its lifecycle is well known (Belharet et al., 2020) and previous studies have already documented the presence of its juveniles in ports, on natural nurseries and on artificial fish nurseries in our study area (Bouchoucha et al., 2016). Current legislation sets the minimum catch size for *D. sargus* at 23cm in the Mediterranean. Generally, this regulation is not respected, particularly by recreational fishing (Font and Lloret, 2014).

## Isis-Fish model

The ISIS-Fish modelling platform (Mahévas and Pelletier, 2004) was used to simulate the life cycle of *D. sargus* exploited by several fishing fleets and test conservation measures on the basis of scenarios. This spatially explicit simulation tool operates through three interconnected sub-models: (i) population dynamics, (ii) fleet dynamics, and (iii) fishery management. The sub models are coupled in space and time, exchanging information in a discrete space and over monthly steps. The first sub-model describes fish population dynamics, accounting for growth, reproduction, mortality, recruitment, and movements between different areas based on the age group of individuals. The second sub-model focuses on fishing activity and describes the spatial dynamics of vessel fishing time as a function of the gear used, the target species, the areas and the fleet to which it belongs. Every month the model simulates species abundance per group, and catches per fleets. Finally, the management sub-model describes the regulations governing the fishing activity based on technical restrictions, spatial limitations of access, catches limitations, etc. (see Mahévas and Pelletier, (2004) and Pelletier et al., (2009) for details). Therefore, ISIS-Fish can describe a complex system with many species, many fishing activities and spatial interactions, as well as the population dynamics of a single species exploited by several fishing activities, as was the case here. This flexibility means that the complexity of the description of the system studied can be adapted to the question raised and the available data. Hereafter, we describe the parametrization of the case study, the calibration process and the sensitivity analyses carried out to build an operational modelling tool and test the effectiveness of several management scenarios.

## Population dynamics

A stage-structured model described the juvenile life stages of *D. sargus* and their movements to adult areas.. The population was partitioned into 15 stages ($C_0$ to $C_{14}$) (Belharet et al., 2020). $C_0$ (the juveniles) includes individuals aged from 1 to 7 months, and stage 1 includes individuals aged from 7 months to 1.5 years. Subsequent stages (2 to 13) each represent a one-year age interval. Individuals in $C_{14}$ are 14.5 years and older. $C_0$ individuals settle in artificial or port nurseries, inside and outside port areas, respectively. The settling capacity is fixed at 10 ind.m$^{-1}$, coupled with the length of each nursery (Doherty, 1991). Each nursery was assigned with a juvenile mortality rate of 80.8% for natural nurseries (Belharet et al., 2020) and 99% for port ones (high estimate, see S1) (Fig. 2). After a period of 6 months, the surviving $C_0$ juveniles leave the nurseries to enter the adult living zone as $C_1$. Each year, the survivors of the stages progress to the higher stages until reaching $C_{14}$. The adult population ($C_1$-$C_{14}$) and its recruitment (*i.e.* moving from $C_0$ to $C_1$) are thus limited by both the surface area of the juvenile habitat and the annual juvenile mortality rate. In the model, the distribution of individuals in the different living zones (nurseries and adult areas) is uniform. All the parameters that shape the life of *D. sargus* in the model were sourced from the literature and are extensively described in SI.1.

## Fishing activities

We distinguished two types of fishing activities: professional and recreational fishing. Five professional fishing gears (long lining, trap, gillnet, trammel net and gangui) and two recreational fishing gears (angling and spearfishing) (Table 1) catch about 17.3 t.year$^{-1}$ (Système d'Information Halieutique, 2022) and 12.4t.year$^{-1}$ of *D. sargus*, respectively (BVA, 2009; Cadiou et al., 2009). Fishing activity was homogeneously distributed throughout the adult life zone of *D. sargus*. Catches were calculated as a function of time, available *D. sargus* biomass and gear technical parameters describing their ability to catch fish (*i.e.* standardisation, selectivity and accessibility, see SI.1 for details).

## Calibration

A simulation consists of running ISIS-Fish over a defined period with a value for each parameter and the initial population abundance, resulting in monthly calculations of population abundance by age stages and catches. The period was set to 15 years to ensure the final population size reflects equilibrium. As the fishing parameters mentioned before lack literature values, they were calibrated to meet two objectives: (i) to align with observations or estimates of catches for each fishing gear (Table 1), and (ii) to correspond to the proportion of catches under the regulatory catch limit size of 23 cm for *D. sargus*. In France, professional and recreational fishers catch 19% and 81% of fish respectively under the regulatory catch limit size (Font and Lloret, 2014), despite the prohibition of capture.

| Gears | Fishing activity | Mean annual catches 2019-2020 (kg.year$^{-1}$) | Source of the data | Mean annual catches of the model after calibration (kg.year$^{-1}$) |
|---|---|---|---|---|
| Long lining | Professional | 14 280 | Declarative data | 14 272 |
| Trap | Professional | 25 | Declarative data | 33 |
| Gillnet | Professional | 780 | Declarative data | 778 |
| Trammel net | Professional | 1 920 | Declarative data | 1423 |
| Gangui (Trawl) | Professional | 300 | Declarative data | 304 |
| Angling | Recreational | 6 000 | Estimations (Cadiou et al., 2009; BVA, 2009) | 6019 |
| Spearfishing | Recreational | 6 400 | Estimations (Cadiou et al., 2009; BVA, 2009) | 6426 |

*Table 1: Annual catches of professional and recreational fishing activities described in the model. Professional catches are calculated from declarative data (REF) while recreational catches are estimates from BVA, (2009); Cadiou et al., (2009).*

Calibration involves running simulations with varied fishing parameter values, selecting the combination that best aligns with objectives (i) and (ii) (see SI.1 for details).

## Management scenarios

The effectiveness of four management scenarios for the *D. sargus* population and catches was tested and compared to a reference scenario (S0) which simulates the current wild population of *D. sargus* (Fig. 2). S0 is used as an initial state from which the conservation scenarios were tested. Scenarios 1 and 2 (S1 and S2) involved rehabilitating 10% (6 006m including 2 477 m of docks already hosting juveniles) and 100% of the docks in the port areas (60 063m including all the docks already hosting juveniles) of the study zone respectively (see SI.2 and Fig. 2). The model incorporates efforts to rehabilitate nursery functions by converting port nurseries and docks into natural nurseries. This involves a reduction in the juvenile mortality rate for already existing port nurseries and the creation of additional natural nurseries for the rest of the port (Fig. 2). These scenarios represent the potential effect of implementing artificial fish nurseries or integrating eco-design in port construction. Scenario 3 (S3) tested the strict application of the already existing regulations on the *D. sargus* catch size alone (strengthening of fisheries controls). Scenario 4 (S4) is a combination of S1 and S3 (see SI.2 and Fig. 2).

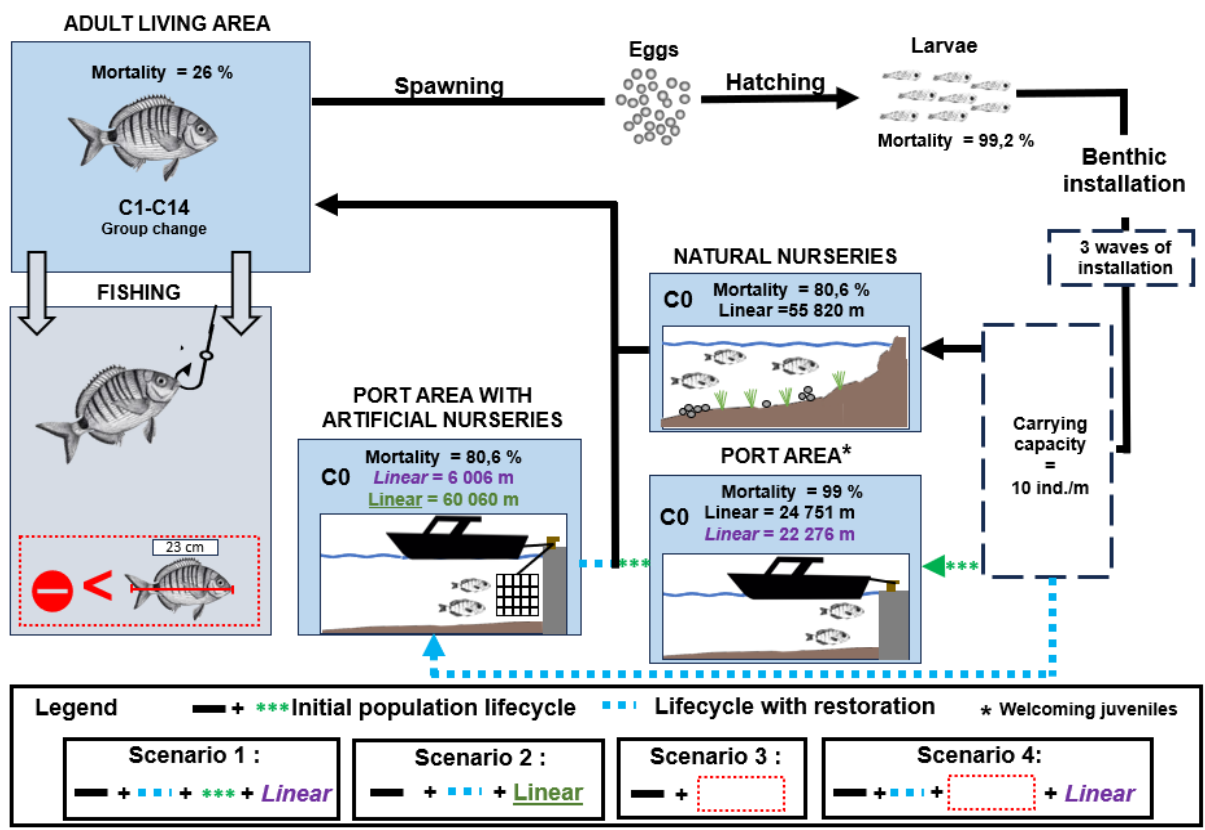

*Figure 2: Parameterization of the D. sargus life cycle model and impacts on the life cycle of D. sargus of conservation measures through scenarios. S1: rehabilitating the nursery function of 10% of the docks. S2: rehabilitating the nursery function of 100% of the docks. S3: banning fishing under 23cm in length, S4: application of S1 and S3 together.*

The effectiveness of each scenario was assessed based on the change in the *D. sargus* $C_1$-$C_{14}$ population size (fish over 8cm), as well as on catches, compared to those in S0.

# Uncertainty around parameter values and sensitivity analysis

There is uncertainty in the literature regarding the settlement capacity of nurseries (including ports and natural nurseries), the number of arrivals of larvae to coastal habitat per year, and the mortality rates of *D. sargus'* life stages (Cuadros et al., 2018). Nevertheless, the *D. sargus* population is constrained by these parameters in the model. Therefore, a range of each parameter variation was defined and an uncertainty analysis was conducted using a simulation design employing random Latin Hypercube Sampling (LHS). LHS is a stratified Monte Carlo sampling approach within the ranges of uncertainties, assuming equal probability for all values in the range. This approach provides a population abundance range for each scenario, offering insights into the potential variability in outcomes according to the variability in parameter values (see SI.3). Here, 5 000 simulations were run for each scenario, each corresponding to a unique combination of values for the parameters. The mean gains or losses associated with a standard deviation for each scenario over the 5 000 simulations was calculated compared with S0 so that the result is given in proportion to the result simulated by S0, for abundance $N_a$ in January (equation 1), mean weight change $W_i$ per individual in January, (equation 2) and catches of professional $L_p$ and recreational $L_r$ fishing (equations 3 & 4), respectively:

$$N_a = \frac{1}{5000} \sum_{k=1}^{5000} \left( \frac{\sum_{n=1}^{14} c_{n_0}}{\sum_{n=1}^{14} c_{n_s}} - 1 \right) * 100 \quad (eqn\ 1)$$

$$W_i = \frac{1}{5000} \left( \sum_{k=1}^{5000} \left( \frac{\sum_{n=1}^{14} wc_{n_s} * \sum_{n=1}^{14} c_{n0}}{\sum_{n=1}^{14} c_{n_s} * \sum_{n=1}^{14} wc_{n0}} - 1 \right) * 100 \quad (eqn\ 2) \right.$$

$$L_p = \frac{1}{5000} \sum_{k=1}^{5000} \left( \frac{\sum_{m_p=1}^{5} Ds_{m_p}}{\sum_{m_p=1}^{5} Do_{m_p}} - 1 \right) * 100 \quad (eqn\ 3)$$

$$L_r = \frac{1}{5000} \sum_{k=1}^{5000} \left( \frac{\sum_{m_r=1}^{5} Ds_{m_r}}{\sum_{m_r=1}^{5} Do_{m_r}} - 1 \right) * 100 \quad (eqn\ 4)$$

With $c_{n_0}$ being the number of individuals in C$_n$ of S0 (0) in January, $c_{n_s}$ the number of individuals in stage $n$ and $wc_{n_s}$ and $wc_{n_0}$ the total weight of individuals in stage n, of scenario $s$ or 0 (with $s \in$ {1,2,3,4} in January. And for ii and iii, $Ds_{m_p}$ and $Ds_{m_r}$ the total annual catches of the gear $m_p$ of simulated professional and recreational fishing respectively, $Do_{m_p}$ and $Do_{m_r}$ the total annual catches of gear $m_p$ of professional and recreational fishing observed respectively.

## Results

### Reference scenario (S0)

The 5 000 simulations for S0 of *D. sargus* gave an equilibrium population in January of 1.26 x 10$^6$ ± 0.66 x 10$^6$ individuals (mean ± SD). This population was renewed annually in September by 415.1 x 10$^3$ ± 204.1 x 10$^3$ recruits, including 352.6 x 10$^3$ ± 165.6 x 10$^3$ individuals from natural nurseries and 62.5 x 10$^3$ ± 44.5 x 10$^3$ individuals from port areas. Each year 9.4 ± 5.2 % of individuals over 8 cm were harvested by the fishing activities.

### Management scenarios

The implementation of the scenarios led to a significant increase in the adult population of D. sargus, visible from the first years after installing artificial habitats and contrasted gains in catches between recreastional and profesionnal fleets.

The population approached equilibrium after 7 years and reached it by the 15th year, following a logarithmic pattern (Fig. 3a). Even with simulated uncertainty around the parameter values relating to settlement capacity, the annual volume of settlers arrived, and mortality rate, the ranking between scenarios was not modified. S1 led to an increase of 7.97 ± 1.35% in the adult population. This gain is 79.53 ± 13.48%, 16.83 ± 0.53% and 26.24 ± 1.20% for Scenarios 2, 3 and 4 respectively (Fig. 3b). This gain is uniform for classes C1 to C14 in S1 and S2. In contrast, for S3 and S4, the abundances of classes C5 to C14 increased by 34.84 ± 0.03% and 45.58 ± 1.82%, respectively, compared with gains ranging from 2.46 ± 0.02% and 10.62 ± 2.40% to 28.17 ± 0.02% and 38.38 ± 1.75% for classes C1 to C4 (See SI.4). The ranking of gains remains consistent across classes, except for $C_1$, where the gains are greater following the application of S1 compared to S3 (See SI.4).

The average mass per individual of the populations resulting from S1 and S2 remains identical to that of the initial population. Conversely, S3 and S4 lead to an increase in the average mass per individual of 7.79 ± 0.04%.

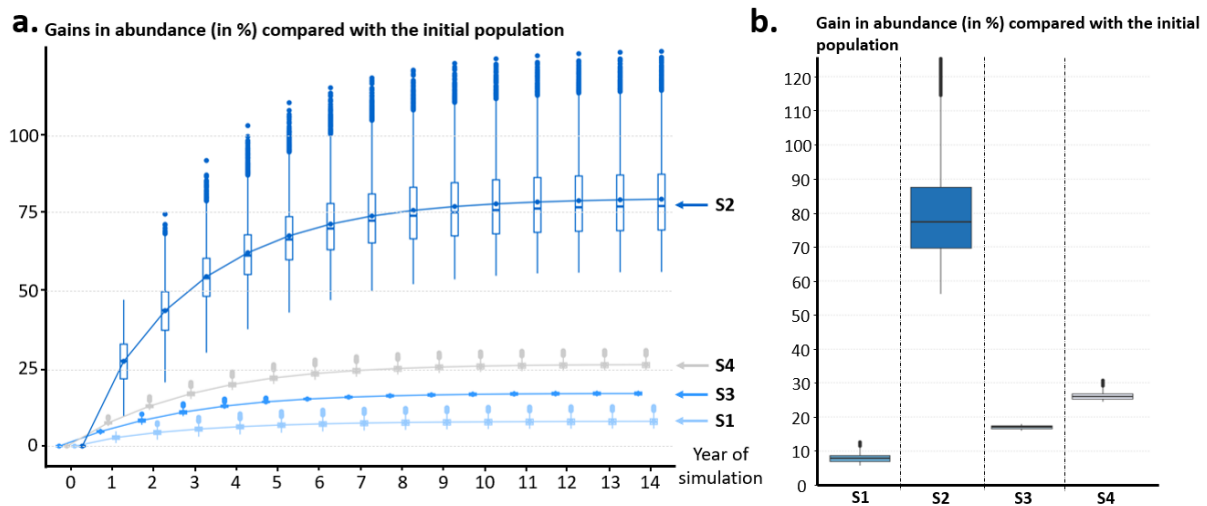

*Figure 3: Gains in abundance (in %) of adult D. sargus individuals (in January) as a function of the scenario implemented. A.) Boxplot of the yearly gain in abundance (the line link the mean of the gains for each boxplot), B.) Boxplot of the total abundance of the population after 15 years. Boxplots present the median at their centre surrounded by the first and third quantiles. The outliers are represented by black dots.*

S1 results in an increase in catches by professional and recreational fishers of 7.97 ± 1.35%. It rises to 79.47 ± 13.48% for S2. While S3 and S4 result in an increase in catches from professional fishing of 12.12 ± 1.8% and 21.03 ± 1.22% respectively, they lead to a decrease in catches from recreational fishing of -70.42 ± 2.21% and –68.1 ± 2.1% (Fig. 4). For both professional and recreational fishing, these percentages are identical for all gears (See SI.4).

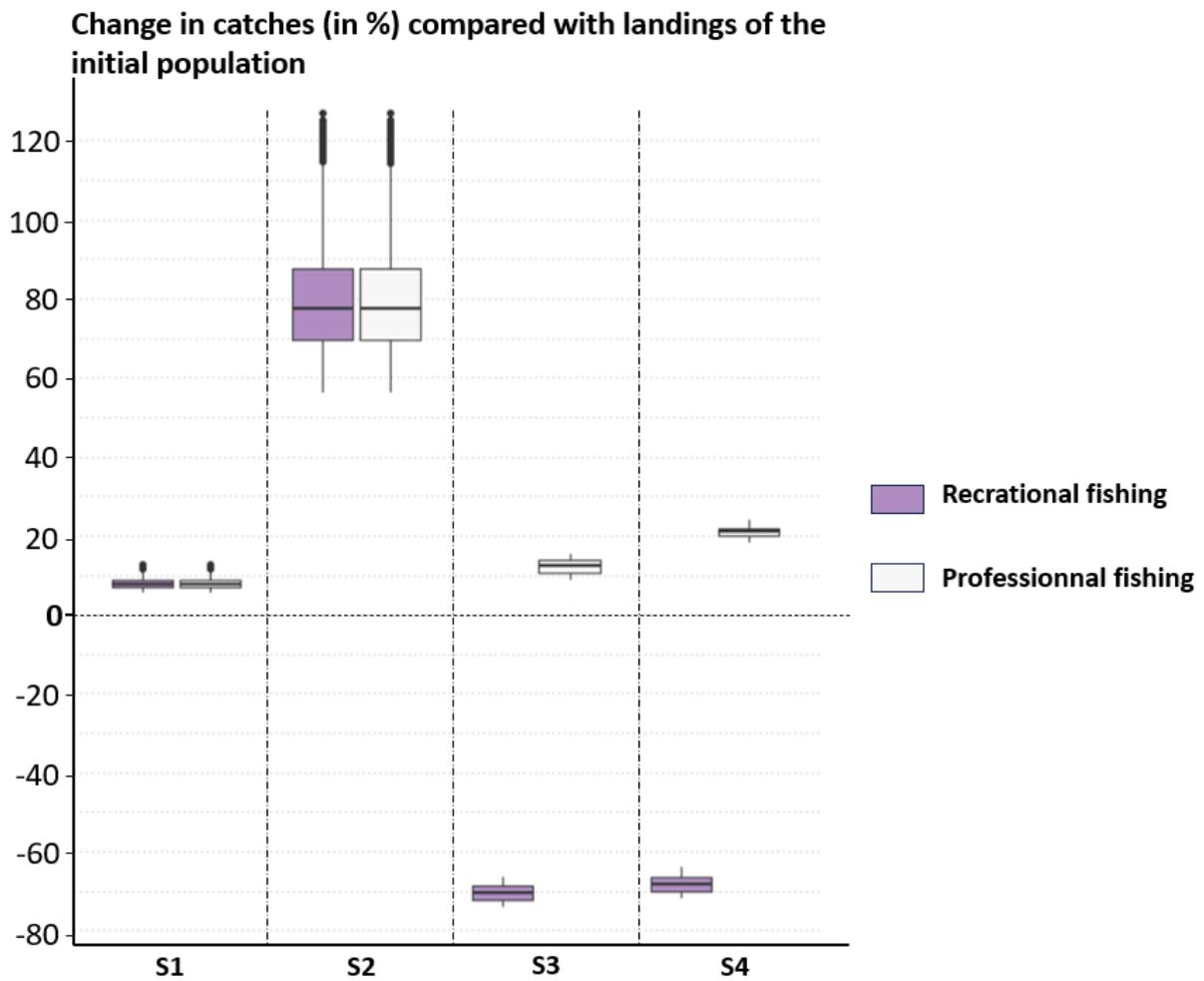

*Figure 4: Boxplot of the effect of the different scenarios on professional (grey) and recreational (violet) fishing annual catches at equilibrium, and expressed as a gain or loss (in %). Boxplots represent the median at their centre surrounded by the first and third quantiles. The outliers are represented by black dots.*

# Discussion

To the best of our knowledge, this study represents the first quantitative assessment of fish nursery rehabilitation projects' effectiveness in port areas, aimed at supporting coastal fish populations and fisheries. In addition, this work facilitated a comparison of the effectiveness between rehabilitation actions and a fishing control measure. Although small-scale rehabilitation projects demonstrate promise in supporting fish populations and fisheries, their efficacy remains limited compared to fishing management measures. Moreover, our findings underscore the greater effectiveness of employing multiple conservation measures simultaneously rather than implementing them individually, providing valuable insights for environmental stakeholders in managing marine coastal areas.

Our findings reveal that the rehabilitation of fish nurseries in the Toulon port area has limited population level impact when 6km of port are artificially equipped, and a significant restoration effect when 100% of the port is equipped.

These results suggest that current port nursery rehabilitation projects may be relatively small in scale, potentially limiting their real impact on fish populations. Indeed, existing studies on rehabilitation projects in port areas have reported rehabilitated surface areas of 150m$^2$ (15m linear)(Joubert et al., 2023), 24m$^2$ (20m linear) across five ports (Bouchoucha et al., 2016), 4m$^3$ across 6m linear (Patranella et al., 2017). These figures fall significantly short of the 6km of equipped docks in our scenario 1 (S1)that leads to a 7.97% ± 1.35% increase in the adult *D. sargus* population. To significantly impact coastal fish populations and fisheries, it is therefore crucial to reconsider and potentially increase the spatial extent of rehabilitation projects. This critique aligns with common consideration for restoration projects, particularly in the marine environment (Airoldi et al., 2021; Chapman et al., 2018). However, it is essential to note that, beyond financial considerations, project sizes cannot be expanded infinitely, and socio-ecological gains eventually reach a plateau. In our study, the gain between S1 and S2 should be regarded as an unattainable maximum benefit. In the model, the adult habitat have not been limited as the densities computed still lower than the one currently observed in certain protected area (Belharet et al., 2020) inducing in linear results the raise in percentage of the abundance of *D. sargus* when S1 and S2 are deployed. The model functions on the assumption that each linear meter of artificial fish nurseries in ports equals one additional meter of natural nursery. It implies that recruitment to the adult population is solely limited by the surface area of the juvenile habitat while the number of settling larvae is never limiting. This hypothesis also assumes that the observed higher density of juvenile fish on artificial fish nurseries than on bare docks (Bouchoucha et al., 2016; Joubert et al., 2023) is exclusively linked to reduced juvenile mortality and not to the attraction of juveniles already present on neighbouring infrastructures. However, this this aspect remains unclear and is a topic of debate concerning the local effectiveness of artificial structures in general (Grossman et al., 1997). This also suggests that the excess mortality of juveniles in port areas compared to natural nurseries is solely due to the lack of habitat complexity. However although the growth and condition of juveniles are sometimes equivalent inside and outside a port for some *Sparidae* species (Bouchoucha et al., 2018), juvenile exposure to contaminants can have long-term effects and lead to a reduction in recruitment success (Kazour et al., 2020), and could lead to a drop in individual fitness during adult life (Bouchoucha et al., 2018). These assumptions likely lead our model to overestimate the effectiveness of artificial fish nurseries in port areas, with actual gains from projects likely to be half of our estimates. The gains in terms of fish population and catches for very small projects are almost certainly negligible. Adding artificial nurseries on port docks at the dimensions outlined in S1, S2, and S4 might be deemed prohibitively expensive and unrealistic. However, a more feasible approach could involve designing docks to be eco-friendly during port construction and incorporating greater structural complexity. This approach is in line with the linear objectives tested in S1 and S2, presenting a realistic and cost-effective option (Airoldi et al., 2021).

Our results suggest that rehabilitating nursery functions in port areas is less effective compared to fishing management measures. This study confirms that while habitat loss significantly impacts coastal fish populations, for a commercial species such as *D. sargus*, fishing remains the main pressure constraining population growth (Yan et al., 2021). Therefore, fishing pressure currently limits the effectiveness of rehabilitation projects. This pressure is particularly significant, as individuals below the minimum catch size (23cm) are heavily targeted (Tsikliras and Stergiou, 2014), despite regulations in France prohibiting their capture. To support the renewal of fish populations, the foremost priority would be to regulate fishing activities (or apply existing regulations), especially recreational fishing for coastal species, before considering rehabilitation projects. This aligns with the Society for Ecological Restoration's recommendations to reduce the main pressures causing degradation before initiating restoration activities. (Gann et al., 2019). Nevertheless, the attractiveness of rehabilitation increases when combined with regulation measures. In our study, S4, combining reduced fishing pressure with artificial habitats in ports, provides a gain over 1.44% higher than the sum of gains in S1 and S3. The management measures in these scenarios target different segments of the population: S1 and S2 directly impact $C_0$, while S3 impacts classes $C_1$ to $C_4$. Gains from S1 and S2 uniformly affect classes $C_1$ to $C_{14}$, whereas gains from S3 increasingly affect classes $C_1$ to $C_4$ and then uniformly affect classes $C_5$ to $C_{14}$.

Therefore, rehabilitation projects should never be considered as a substitute for regulations or protection measures but can be regarded as a complementary measure, leading to a synergistic effect (Possingham et al., 2015). When no other viable options exist, they can also be regarded as mitigation projects. The gains are minimal, but they are still preferable to taking no action. A recent study highlighted the greater potential for restoring estuarine nurseries in the north of France than our study around Toulon (Gernez et al., 2023). The Mediterranean lacks such intensive fisheries controls as the French Exclusive Economic Zones of the Atlantic or the Channel, particularly for inshore and recreational fishing (Cardinale et al., 2017). This comparison supports the fact that the implementation of conservation measures must be studied on a case-by-case basis, as regional context greatly influences their success.

The gains associated with each scenario's impact on *D. sargus* abundance are assessed after 15 years, representing a new demographic balance. This balance reflects the replacement of all individuals in the initial population with those born after the scenario's implementation. Introducing natural stochastic phenomena to the lifecycle of *D. sargus* would extend timescales and alter the logarithmic pattern of fish abundance (Hastings et al., 2021).

For environmental managers, this implies that gains in population abundance during the initial years after implementing management measures may be lower than those observed in this study. Therefore, these solutions require long-term planning, with objectives spanning over a decade, regardless of the chosen conservation measure (Airoldi et al., 2021).

While models are valuable, they involve several approximations. Therefore, the results should not be seen as an exact reproduction of reality, and recommendations should be approached with caution. However, the flexible nature of the ISIS-Fish model allows for refinement, such as considering fishers' behaviour and compliance with regulation.

In our work catches by recreational fishers are strongly affected by scenarios 3 and 4. In reality, fishers will likely adapt by targeting larger individuals, tempering the decrease in recreational fishing catches. This activity, important for the local economy, may be less impacted than predicted. In this case, additional data, regarding the spatialization of fishing activity and fish distribution, as well as testing conservation measures on different species and scales, can lead to more precise recommendations. Management measures should align with the expected objectives. To assess the effectiveness of artificial nurseries in port areas, we focused on adult fish population abundance and fishery catches.. Other studies evaluated different indicators for alternative perspectives on project success. For instance, if the goal is to enhance the local ecological status of port areas, then a rehabilitation project should be considered (Bouchoucha et al., 2016; Joubert et al., 2023). The societal impact of this kind of small-scale project, in terms of communication and community commitment to the environmental cause, should not be underestimated. Moreover, pilot projects are essential before large-scale investments (Airoldi et al., 2021). Every human project has its drawbacks and if artificial nursery projects are implemented in port areas, precautions must be taken to avoid additional pollution (Cooke et al., 2023) and the introduction of potentially invasive non-indigenous species (Gauff et al., 2023).

## Author contribution

Etienne Joubert: Conceptualization, Methodology, Formal analysis, Writing – Original Draft; Charlotte Sève and Stéphanie Mahévas: Methodology, Resources, Writing – Review & Editing; Adrian Bach: Conceptualization, Review & Editing; Marc Bouchoucha: Conceptualization, Methodology, Writing – Review & Editing, Project administration, Funding acquisition.

# Acknowledgement

This study was supported, by the Direction Interrégionale de la Mer Mediterrannée (DIRM) and the Horizon Europe CLIMAREST Project (Coastal Climate Resilience and Marine Restoration Tools for the Arctic Atlantic basin) (GA no. 101093865) We are grateful to the Seaboost company for having financed Etienne Joubert's work during the third year of this project. In addition, we wish to acknowledge the anonymous reviewers for their valuable comments and suggestions that improved our manuscript.

# Conflicts of interest

The Seaboost company, which financed Etienne Joubert work during the third year of this project, is specialized in building and selling artificial structures aiming to rehabilitate the fish nursery function in port areas. There are no other conflicts of interest to declare.

# Data availability statement

Data available via the Seanoe Repository :

# Supporting information

Supporting information about the method and the model are available at …

# References

Airoldi, L., Beck, M., Firth, L., Bugnot, A.B., Steinberg, P., Dafforn, K., 2021. Emerging solutions to return nature to the urban ocean. Annu. Rev. Mar. Sci. 13, 445–477. https://doi.org/10.1146/annurev-marine-032020-020015.

Bach, A., Minderman, J., Bunnefeld, N., Mill, A.C., Duthie, A.B., 2022. Intervene or wait? A model evaluating the timing of intervention in conservation conflicts adaptive management under uncertainty. Ecol. Soc. 27. https://doi.org/10.5751/ES-13341-270303

Beck, M.W., Heck, K.L., Able, K.W., Childers, D.L., Eggleston, D.B., Gillanders, B.M., Halpern, B., Hays, C.G., Hoshino, K., Minello, T.J., Orth, R.J., Sheridan, P.F., Weinstein, M.P., 2001. The identification, conservation, and management of estuarine and marine nurseries for fish and invertebrates: A better understanding of the habitats that serve as nurseries for marine species and the factors that create site-specific variability in nursery quality will improve conservation and management of these areas. BioScience 51, 633–641. https://doi.org/10.1641/0006-3568(2001)051[0633:TICAMO]2.0.CO;2

Belharet, M., Franco, A.D., Calò, A., Mari, L., Claudet, J., Casagrandi, R., Gatto, M., Lloret, J., Sève, C., Guidetti, P., Melià, P., 2020. Extending full protection inside existing marine protected areas, or reducing fishing effort outside, can reconcile conservation and fisheries goals. J. Appl. Ecol. 57, 1948–1957. https://doi.org/10.1111/1365-2664.13688

Bouchoucha, M., Brach-Papa, C., Gonzalez, J.-L., Lenfant, P., Darnaude, A.M., 2018. Growth, condition and metal concentration in juveniles of two Diplodus species in ports. Mar. Pollut. Bull. 126, 31–42. https://doi.org/10.1016/j.marpolbul.2017.10.086

Bouchoucha, M., Darnaude, A.M., Gudefin, A., Neveu, R., Verdoit-Jarraya, M., Boissery, P., Lenfant, P., 2016. Potential use of marinas as nursery grounds by rocky fishes: insights from four Diplodus species in the Mediterranean. Mar. Ecol. Prog. Ser. 547, 193–209. https://doi.org/10.3354/meps11641

BVA, 2009. Enquête relative à la pêche de loisir (récréative et sportive) en mer en Métropole et dans les DOM. Synthèse des résultats finaux 1–13.

Cadiou, G., Boudouresque, C.F., Bonhomme, P., Le Diréach, L., 2009. The management of artisanal fishing within the Marine Protected Area of the Port-Cros National Park (northwest Mediterranean Sea): a success story? ICES J. Mar. Sci. 66, 41–49. https://doi.org/10.1093/icesjms/fsn188

Cardinale, M., Osio, G.C., Scarcella, G., 2017. Mediterranean Sea: A Failure of the European Fisheries Management System. Front. Mar. Sci. 4.

Chapman, M.G., Underwood, A.J., Browne, M.A., 2018. An assessment of the current usage of ecological engineering and reconciliation ecology in managing alterations to habitats in urban estuaries. Ecol. Eng. 120, 560–573. https://doi.org/10.1016/j.ecoleng.2017.06.050

Connell, S.D., Jones, G.P., 1991. The influence of habitat complexity on postrecruitment processes in a temperate reef fish population. J. Exp. Mar. Biol. Ecol. 151, 271–294. https://doi.org/10.1016/0022-0981(91)90129-K

Cooke, S.J., Piczak, M.L., Vermaire, J.C., Kirkwood, A.E., 2023. On the troubling use of plastic 'habitat' structures for fish in freshwater ecosystems – or – when restoration is just littering. FACETS 8, 1–19. https://doi.org/10.1139/facets-2022-0210

Cuadros, A., Basterretxea, G., Cardona, L., Cheminée, A., Hidalgo, M., Moranta, J., 2018. Settlement and post-settlement survival rates of the white seabream (Diplodus sargus) in the western Mediterranean Sea. PLOS ONE 13, e0190278. https://doi.org/10.1371/journal.pone.0190278

De Vos, J.M., Joppa, L.N., Gittleman, J.L., Stephens, P.R., Pimm, S.L., 2015. Estimating the normal background rate of species extinction. Conserv. Biol. 29, 452–462. https://doi.org/10.1111/cobi.12380

DeAngelis, D.L., Diaz, S.G., 2019. Decision-Making in Agent-Based Modeling: A Current Review and Future Prospectus. Front. Ecol. Evol. 6. https://doi.org/10.3389/fevo.2018.00237

Doherty, P.J., 1991. Spatial and temporal patterns in recruitment. Ecol. Fishes Coral Reefs 509, 261–293. https://doi.org/10.1016/B978-0-08-092551-6.50015-5

Font, T., Lloret, J., 2014. Biological and Ecological Impacts Derived from Recreational Fishing in Mediterranean Coastal Areas. Rev. Fish. Sci. Aquac. 22, 73–85. https://doi.org/10.1080/10641262.2013.823907

Gann, G.D., McDonald, T., Walder, B., Aronson, J., Nelson, C.R., Jonson, J., Hallett, J.G., Eisenberg, C., Guariguata, M.R., Liu, J., Hua, F., Echeverría, C., Gonzales, E., Shaw, N., Decleer, K., Dixon, K., 2019. International principles and standards for the practice of ecological restoration. Second edition. Restor. Ecol. 27, S1–S46. https://doi.org/10.1111/rec.13035

Gauff, R.P.M., Joubert, E., Curd, A., Carlier, A., Chavanon, F., Ravel, C., Bouchoucha, M., 2023. The elephant in the room: Introduced species also profit from refuge creation by artificial fish habitats. Mar. Environ. Res. 185, 105859. https://doi.org/10.1016/j.marenvres.2022.105859

Gernez, M., Champagnat, J., Rivot, E., Le Pape, O., 2023. Potential impacts of the restoration of coastal and estuarine nurseries on the stock dynamics of fisheries species. Estuar. Coast. Shelf Sci. 108557. https://doi.org/10.1016/j.ecss.2023.108557

Glaum, P., Cocco, V., Valdovinos, F.S., 2020. Integrating economic dynamics into ecological networks: The case of fishery sustainability. Sci. Adv. 6, eaaz4891. https://doi.org/10.1126/sciadv.aaz4891

Grossman, G.D., Jones, G.P., Seaman, W.J., 1997. Do Artificial Reefs Increase Regional Fish Production? A Review of Existing Data. Fisheries 22, 17–23. https://doi.org/10.1577/1548-8446(1997)022<0017:DARIRF>2.0.CO;2

Harmelin-Vivien, M.L., Harmelin, J.G., Leboulleux, V., 1995. Microhabitat requirements for settlement of juvenile sparid fishes on Mediterranean rocky shores. Hydrobiologia 300, 309–320. https://doi.org/10.1007/BF00024471

Hastings, A., Abbott, K.C., Cuddington, K., Francis, T.B., Lai, Y.-C., Morozov, A., Petrovskii, S., Zeeman, M.L., 2021. Effects of stochasticity on the length and behaviour of ecological transients. J. R. Soc. Interface 18, 20210257. https://doi.org/10.1098/rsif.2021.0257

Joubert, E., Gauff, R.P.M., de Vogüé, B., Chavanon, F., Ravel, C., Bouchoucha, M., 2023. Artificial fish nurseries can restore certain nursery characteristics in marine urban habitats. Mar. Environ. Res. 106108. https://doi.org/10.1016/j.marenvres.2023.106108

Kazour, M., Jemaa, S., El Rakwe, M., Duflos, G., Hermabassiere, L., Dehaut, A., Le Bihanic, F., Cachot, J., Cornille, V., Rabhi, K., Khalaf, G., Amara, R., 2020. Juvenile fish caging as a tool for assessing microplastics contamination in estuarine fish nursery grounds. Environ. Sci. Pollut. Res. 27, 3548–3559. https://doi.org/10.1007/s11356-018-3345-8

Le Pape, O., Vermard, Y., Guitton, J., Brown, E.J., van de Wolfshaar, K.E., Lipcius, R.N., Støttrup, J.G., Rose, K.A., 2020. The use and performance of survey-based pre-recruit abundance indices for possible inclusion in stock assessments of coastal-dependent species. ICES J. Mar. Sci. 77, 1953–1965. https://doi.org/10.1093/icesjms/fsaa051

Macura, B., Byström, P., Airoldi, L., Eriksson, B.K., Rudstam, L., Støttrup, J.G., 2019. Impact of structural habitat modifications in coastal temperate systems on fish recruitment: a systematic review. Environ. Evid. 8, 14. https://doi.org/10.1186/s13750-019-0157-3

Mahévas, S., Pelletier, D., 2004. ISIS-Fish, a generic and spatially explicit simulation tool for evaluating the impact of management measures on fisheries dynamics. Ecol. Model. 171, 65–84.

Meinesz, A., Lefevre, J.R., Astier, J.M., 1991. Impact of coastal development on the infralittoral zone along the southeastern Mediterranean shore of continental France. Mar. Pollut. Bull., Environmental Management and Appropriate Use of Enclosed Coastal Seas 23, 343–347. https://doi.org/10.1016/0025-326X(91)90698-R

Mooser, A., Anfuso, G., Williams, A.T., Molina, R., Aucelli, P.P.C., 2021. An Innovative Approach to Determine Coastal Scenic Beauty and Sensitivity in a Scenario of Increasing Human Pressure and Natural Impacts due to Climate Change. Water 13, 49. https://doi.org/10.3390/w13010049

Patranella, A., Kilfoyle, K., Pioch, S., Spieler, R.E., 2017. Artificial Reefs as Juvenile Fish Habitat in a Marina. J. Coast. Res. 33, 1341–1351. https://doi.org/10.2112/JCOASTRES-D-16-00145.1

Pelletier, D., Mahevas, S., Drouineau, H., Vermard, Y., Thebaud, O., Guyader, O., Poussin, B., 2009. Evaluation of the bioeconomic sustainability of multi-species multi-fleet fisheries under a wide range of policy options using ISIS-Fish. Ecol. Model. 220, 1013–1033. https://doi.org/10.1016/j.ecolmodel.2009.01.007

Pimm, S.L., Jenkins, C.N., Abell, R., Brooks, T.M., Gittleman, J.L., Joppa, L.N., Raven, P.H., Roberts, C.M., Sexton, J.O., 2014. The biodiversity of species and their rates of extinction, distribution, and protection. Science 344, 1246752. https://doi.org/10.1126/science.1246752

Plagányi, É.E., Punt, A.E., Hillary, R., Morello, E.B., Thébaud, O., Hutton, T., Pillans, R.D., Thorson, J.T., Fulton, E.A., Smith, A.D.M., Smith, F., Bayliss, P., Haywood, M., Lyne, V., Rothlisberg, P.C., 2014. Multispecies fisheries management and conservation: tactical applications using models of intermediate complexity. Fish Fish. 15, 1–22. https://doi.org/10.1111/j.1467-2979.2012.00488.x

Planes, S., Jouvenel, J.-Y., Lenfant, P., 1998. Density dependence in post-recruitment processes of juvenile sparids in the littoral of the Mediterranean sea. Oikos 83, 293–300. https://doi.org/10.2307/3546840

Possingham, H.P., Bode, M., Klein, C.J., 2015. Optimal Conservation Outcomes Require Both Restoration and Protection. PLOS Biol. 13, e1002052. https://doi.org/10.1371/journal.pbio.1002052

Poursanidis, D., Topouzelis, K., Chrysoulakis, N., 2018. Mapping coastal marine habitats and delineating the deep limits of the Neptune's seagrass meadows using very high resolution Earth observation data. Int. J. Remote Sens. 39, 8670–8687. https://doi.org/10.1080/01431161.2018.1490974

Refsgaard, J.C., van der Sluijs, J.P., Højberg, A.L., Vanrolleghem, P.A., 2007. Uncertainty in the environmental modelling process – A framework and guidance. Environ. Model. Softw. 22, 1543–1556. https://doi.org/10.1016/j.envsoft.2007.02.004

Scharf, F.S., Manderson, J.P., Fabrizio, M.C., 2006. The effects of seafloor habitat complexity on survival of juvenile fishes: Species-specific interactions with structural refuge. J. Exp. Mar. Biol. Ecol. 335, 167–176. https://doi.org/10.1016/j.jembe.2006.03.018

Système d'Information Halieutique, 2022. Données de production et d'effort de pêche SACROIS (2012 à 2020). Ifremer SIH. https://doi.org/10.12770/8eac4cd1-a546-445c-b3fa-7ed580333403

Tsikliras, A.C., Stergiou, K.I., 2014. Size at maturity of Mediterranean marine fishes. Rev. Fish Biol. Fish. 24, 219–268. https://doi.org/10.1007/s11160-013-9330-x

Vigliola, L., Harmelin-Vivien, M., Biagi, F., Galzin, R., Garcia-Rubies, A., Harmelin, J., Jouvenel, J., Le Direach-Boursier, L., Macpherson, E., Tunesi, L., 1998. Spatial and temporal patterns of settlement among sparid fishes of the genus Diplodus in the northwestern Mediterranean. Mar. Ecol. Prog. Ser. 168, 45–56.

Ward, D., Melbourne-Thomas, J., Pecl, G.T., Evans, K., Green, M., McCormack, P.C., Novaglio, C., Trebilco, R., Bax, N., Brasier, M.J., Cavan, E.L., Edgar, G., Hunt, H.L., Jansen, J., Jones, R., Lea, M.-A., Makomere, R., Mull, C., Semmens, J.M., Shaw, J., Tinch, D., van Steveninck, T.J., Layton, C., 2022. Safeguarding marine life: conservation of biodiversity and ecosystems. Rev. Fish Biol. Fish. 32, 65–100. https://doi.org/10.1007/s11160-022-09700-3

Wilson, S.K., Depczynski, M., Fulton, C.J., Holmes, T.H., Radford, B.T., Tinkler, P., 2016. Influence of nursery microhabitats on the future abundance of a coral reef fish. Proc. R. Soc. B Biol. Sci. 283, 20160903. https://doi.org/10.1098/rspb.2016.0903

Yan, H.F., Kyne, P.M., Jabado, R.W., Leeney, R.H., Davidson, L.N.K., Derrick, D.H., Finucci, B., Freckleton, R.P., Fordham, S.V., Dulvy, N.K., 2021. Overfishing and habitat loss drive range contraction of iconic marine fishes to near extinction. Sci. Adv. 7, eabb6026. https://doi.org/10.1126/sciadv.abb6026